\documentclass[trackchanges,twocolumn]{aastex62}

\usepackage{amsmath}
\usepackage{units}

\newcommand{\nuclei}[2]{\ensuremath{\mathrm{^{#1}#2}}}

\newcommand{\carbon}[1][12]{\nuclei{#1}{C}}

\newcommand{\oxygen}[1][16]{\nuclei{#1}{O}}
\newcommand{\fluorine}[1][19]{\nuclei{#1}{F}}
\newcommand{\neon}[1][20]{\nuclei{#1}{Ne}}
\newcommand{\sodium}[1][23]{\nuclei{#1}{Na}}
\newcommand{\magnesium}[1][24]{\nuclei{#1}{Mg}}
\newcommand{\aluminum}[1][27]{\nuclei{#1}{Al}}

\newcommand{\MESA}{\texttt{MESA}}

\newcommand{\Msun}{\ensuremath{\mathrm{M}_\odot}}

\newcommand{\gcc}{\ensuremath{\mathrm{g\,cm^{-3}}}} 

\newcommand{\Tc}{\ensuremath{T_{\mathrm{\!c}}}} 

\newcommand{\rhoc}{\ensuremath{\rho_{\mathrm{c}}}} 

\newcommand{\Ye}{\ensuremath{Y_{\mathrm{e}}}} 

\newcommand{\Msunyr}{\ensuremath{\rm \Msun\,yr^{-1}}} 

\newcommand{\logRho}{\ensuremath{\log(\rho/\gcc)}} 
\newcommand{\logRhoc}{\ensuremath{\log(\rhoc/\gcc)}} 

\begin{document}
\title{Residual Carbon in Oxygen-Neon White Dwarfs and its Implications for Accretion-Induced Collapse}

\shorttitle{Carbon in ONe WDs and Implications for AIC} 
\shortauthors{Schwab \& Rocha}

\author[0000-0002-4870-8855]{Josiah Schwab}
\altaffiliation{Hubble Fellow}
\affiliation{Department of Astronomy and Astrophysics, University of California, Santa Cruz, CA 95064, USA}
\correspondingauthor{Josiah Schwab}
\email{jwschwab@ucsc.edu}

\author[0000-0003-4474-6528]{Kyle Akira Rocha}
\affiliation{Department of Astronomy and Astrophysics, University of California, Santa Cruz, CA 95064, USA}
\affiliation{Department of Physics and Astronomy, Northwestern University, 2145 Sheridan Road, Evanston, IL 60208, USA}


\begin{abstract}
  We explore the effects of the residual \carbon[12] present in
  oxygen-neon white dwarfs (ONe WDs) on their evolution towards
  accretion-induced collapse (AIC).  We produce a set of ONe WD models
  using \MESA\ and illustrate how the amount and location of the
  residual carbon depends on the initial mass of the star and
  assumptions about rotation and convective overshooting.  We find a
  wide range of possible \carbon[12] mass fractions roughly ranging
  from 0.1 to 10 per cent.  Convection and thermohaline mixing that
  occurs as the ONe WDs cool leads to nearly homogeneous interior
  compositions by the time that AIC would occur.  We evolve these ONe
  WD models and some toy WD models towards AIC and find that
  regardless of the carbon fraction, the occurrence of Urca-process
  cooling due to \sodium[23] implies that the models are unlikely to
  reach carbon ignition before electron captures on \magnesium[24]
  occur.  Difficulties associated with modeling
  electron-capture-driven convective regions in these ONe WDs prevent
  us from evolving our \MESA\ models all the way to thermonuclear
  oxygen ignition and the onset of collapse.  Thus, firm conclusions
  about the effect of carbon on the final fates of these objects await
  improved modeling.  However, it is clear that the inclusion
  of residual carbon can shift the evolution from that previously
  described in the literature and should be included in future models.
\end{abstract}


\section{Introduction}

Stellar evolution provides a variety of pathways that
produce degenerate cores with oxygen-neon (ONe) compositions and
masses near the Chandrasekhar mass.  When the central density of the
core reaches a critical value $\sim\unit[10^{10}]{\gcc}$,
electron-capture reactions on isotopes such as \magnesium[24] and
\neon[20] set in motion a chain of events that lead to the destruction
of the star \citep{Miyaji1980, Miyaji1987}.  This has generally been
thought to involve the collapse of the core and the formation of a
neutron star (NS).

Single stars with masses in the range $\approx \unit[8-10]{\Msun}$
develop degenerate ONe cores while avoiding oxygen ignition
\citep[e.g.,][]{Nomoto1984a, GarciaBerro1994} and thus have the
potential to produce an electron-capture supernova
\citep[ECSN;][]{Nomoto1987b, Hashimoto1993}.  A related
process can occur in binary systems that produce an ONe WD with a
close companion, such that material is added to the ONe WD
\citep{Nomoto1979a}: this could be a system where a non-degenerate
companion donates material onto the WD or in a double WD system that
merges.  These two binary scenarios, which are in close analogy to the
single and double degenerate scenarios for Type Ia supernovae
\citep[e.g.,][]{Maoz2014}, are referred to as accretion-induced
collapse (AIC) and merger-induced collapse (MIC) respectively.
AIC and MIC, by virtue of operating in old stellar populations, have long
been invoked as a method of producing young NSs in globular clusters
\citep{Lyne1996, Boyles2011}.  Stellar population synthesis
calculations indicate AIC and MIC may occur at $\approx$ 10\% of the
thermonuclear supernova rate \citep{Ruiter2018}.

The final phase of ECSN/AIC/MIC occurs when the density-driven
initiation of exothermic electron-capture reactions causes the
thermonuclear ignition of oxygen fusion at or near the center of the
ONe core.  This leads to the development of an oxygen-burning
deflagration wave that then propagates outward through the star.
Further electron-capture reactions on the deflagration ashes sap the
pressure support.  The final fate of the star is set by a competition
between these two processes.  The timescales over which these
processes occur are sensitive to density, and so the central density
of the ONe core when oxygen ignition occurs has been
understood as the primary determinant of the final fate.

One-dimensional calculations \citep[e.g.,][]{Nomoto1991,
  Gutierrez1996} suggested that this process led to collapse to a
NS. However, it was understood that this conclusion was sensitive to
the presence and efficiency of mixing processes in the star.  Current
models of electron-capture-initiated collapse, which have only
recently begun to harness the power of multidimensional hydrodynamics
codes, are extremely close to the threshold between implosion and
explosion.  \citet{Jones2016} find that in some cases these objects do
not implode, but instead explode, leaving behind a low mass bound
remnant. \citet{Leung2017} vary a variety of model parameters and find
that the outcome switches between explosion and implosion within
existing uncertainties.

Therefore, an important step in understanding the final fate of these objects is
improving the modeling of the phase in which the WD approaches the
Chandrasekhar mass.  Such calculations inform the initial conditions
for the collapse calculations.  Current models of ECSN progenitors
generally use relatively large nuclear networks
\citep[e.g.,][]{Jones2013, Takahashi2013}, and so the compositions at
the time of collapse are self-consistently set by the preceding stellar
evolution.  In contrast, AIC models have primarily used homogeneous
degenerate cores where the focus is restricted to the most abundant
isotopes \oxygen[16], \neon[20], and \magnesium[24]
\citep[e.g.,][]{Canal1992, Gutierrez1996, Schwab2015}.
\citet{Gutierrez2005} extend this to include the effects of
\carbon[12] and \sodium[23], which can be present with mass fractions
at the percent level. \citet{Gutierrez2005} find that \carbon[12] can
lead to low density explosions, but that \sodium[23] has little
effect.  Using a more accurate treatment of the relevant weak reaction
rates (\citealt{Paxton2015}; see also \citealt{Fuller1985,Toki2013,
  MartinezPinedo2014, Suzuki2016}), \citet{Schwab2017a} demonstrate
that  Urca-process cooling  due to \sodium[23] has a significant effect on the thermal state of the WD.

In this paper, we revisit the role of \carbon[12] in the evolution of
massive ONe WDs towards AIC.
In Section~\ref{sec:how-much-C}, we generate a set of massive ONe WD models and characterize their chemical compositions.
In Section~\ref{sec:homogenization}, we describe how the composition profile of the WD changes during its evolution towards collapse.
Guided by these results, in Section~\ref{sec:carbon} we use a set of simple WD
models to demonstrate the effects of \carbon[12].
In Section~\ref{sec:to-collapse}, we evolve our realistic WD models towards collapse.
In Section 6, we conclude and indicate areas of important remaining
uncertainty in their evolution towards collapse.



\section{Composition of Oxygen-Neon WDs from Stellar Models}
\label{sec:how-much-C}

Super asymptotic giant branch (SAGB) stars that produce ONe WDs first produce partially degenerate CO cores.
Off-center carbon ignition
then occurs, leading to the formation of a 
convectively-bounded ``flame'' that propagates inward \citep[e.g.,][]{Timmes1994}.
The ashes of carbon burning are dominated by \oxygen[16] and \neon[20]
with \sodium[23] and \magnesium[24] also being produced at mass
fractions of $\approx 5\%$.  Some of the \carbon[12] can remain
unburned, with the amount and its distribution within the WD varying
with the mass of the star and assumptions related to the propagation
of the flame \citep{Siess2006}.

In order to produce a set of ONe WD models with self-consistent abundance
profiles, we evolve a set of SAGB star models using
Modules for Experiments in Stellar Astrophysics
\citep[\MESA;][]{Paxton2011, Paxton2013, Paxton2015, Paxton2018}.
In
Sections~\ref{sec:mesa-setup} and \ref{sec:mesa-agb-results} we
describe these models and their compositions.  In
Section~\ref{sec:siess-comparison} we compare our results to the
SAGB models of \citet{Siess2007A&A}.
We note that these SAGB models are constructed within a single-star framework,
  though ONe WDs that undergo AIC necessarily have binary companions.  Exploring
  how the binary evolution (though mass transfer, tides, etc.) affects the properties of the WDs
  that we expect to undergo AIC is an interesting avenue for future work.

\subsection{Setup of SAGB \MESA\ models}
\label{sec:mesa-setup}

\cite{Farmer2015ApJ} use \MESA\ to investigate the properties of
carbon burning in SAGB stars as a function of the initial stellar mass
and rotation rate as well as mixing parameters including convective
overshooting.  They use fine spatial and temporal resolution in
order to carefully resolve the carbon flame.  We therefore elect to use
the \MESA\ input files from \citet{Farmer2015ApJ} as the framework for
producing ONe WDs with realistic composition profiles.  We updated the
\citet{Farmer2015ApJ} input files to be compatible with \MESA\ version
9793.  We then evolve models from the pre-main sequence to the SAGB
and through the carbon flame phase.

We now summarize some of the key input physics and modeling
assumptions made by \cite{Farmer2015ApJ} and hence adopted in our
models.  For complete details, we refer the reader to
\cite{Farmer2015ApJ} and the \MESA\ instrument papers
\citep{Paxton2011, Paxton2013, Paxton2015, Paxton2018}.

We use the \MESA\ nuclear network \texttt{sagb\_NeNa\_MgAl.net} which
uses 22 isotopes to cover the H, He, and C burning phases (including
the pp chains and the CNO, NeNa, and MgAl cycles).
%
Mass loss is treated via the Reimers mass loss prescription
\citep{Reimers1975} on the RGB with $\eta = 0.5$ and a Bl{\"o}cker
mass loss prescription \citep{Bloecker1995a} on the AGB with
$\eta = 0.05$.
%
Convective stability is evaluated using the Ledoux criterion.
Semiconvection is included via the prescription of \citet{Langer1985}
with a dimensionless efficiency factor of 0.01.  Thermohaline mixing
is included using the prescription of \citet{Kippenhahn1980a} with a
dimensionless efficiency factor of 1.
Convective overshooting is implemented in \MESA\ via an exponential
decay of the convective mixing diffusion coefficient beyond the fully
convective boundary \citep{Herwig2000}.  The lengthscale of this
decay, and hence the extent of the overshooting region, is controlled
by the dimensionless parameter $f_{\rm ov}$ which multiplies the local
pressure scale height.  When we include overshooting, we use an
overshooting efficiency of $f_{\rm ov} = 0.016$.
%
Rotation in \MESA\ uses the shellular approximation \citep{Meynet1997}
and is described in \cite{Paxton2013}.  Stellar models are given an
initial solid body rotation rate at ZAMS.   When we include rotation in
our models, we use an initial rotation rate of 0.25 the critical
rotation rate $\omega_{\rm crit}$. 
%
The \MESA\ implementation of rotation and angular momentum transport
closely follows that of \citet{Heger2000a} and \citet{Heger2005}.  Transport is
treated within the diffusion approximation and considers the effects
of the following processes: the dynamical shear instability (DSI),
secular shear instability (SSI), Eddington–Sweet circulation (ES),
Goldreich–Schubert–Fricke instability (GSF), and Spruit–Tayler dynamo
(ST).
We do not include the effects of rotationally-enhanced mass loss or
other angular momentum loss mechanisms such as magnetic braking.
While many of these choices have associated uncertainties and caveats,
on the whole, we believe that this produces models of SAGB stars that
are broadly representative of those in the literature.

Following SAGB stars through their final phases is computationally 
demanding.  For a recent review of some of the challenges of 
modeling stars in this mass range see \cite{Doherty2017PASA}. 
We invest the effort to carefully track the off-center carbon
burning as it migrates to the center and forms the degenerate ONe
core.  However, difficult evolutionary phases remain before the star
leaves behind a WD.
One challenge comes from needing to track a large number of thermal
pulses (hundreds to thousands).  This would require a prohibitively
large number of timesteps, so we wish to halt our models before the
TP-SAGB phase.\footnote{Stopping before the thermal pulses (as opposed
  to stopping after a few thermal pulses) also proved to make it
  simpler to artificially remove the envelope.}  This does mean that
the WD models that we produce have somewhat different masses than if
they had been allowed to evolve though the TP-SAGB.
Another challenge comes from contending with those models that
experience ``dredge-out'' events.  This occurs when the hydrogen-rich
convective envelope merges with the helium-burning convective shell
\citep[e.g.,][]{Ritossa1999, GilPons2010}.  This leads to a hydrogen
flash as protons are mixed into regions of high temperature on the
convective turnover time.  During this phase, our \MESA\ models
encounter numerical difficulties and are unable to continue.  Stellar
evolution codes may be poor tools to follow these dredge-out events
as the hydrogen burning can become sufficiently rapid that standard
assumptions about mixing length theory may be unlikely to hold
\citep{Jones2016a}.

In summary, we cannot simply allow our \MESA\ SAGB models to naturally
evolve, eject their envelopes, and produce WDs.  Since the challenges
occur after completion of the carbon flame phase (and thus after the
core composition is done evolving), we decide to manually halt the
evolution and eject the envelope.\footnote{We note that in many evolutionary
scenarios leading to AIC, the envelopes of the progenitor stars have
already been stripped through binary interaction and thus never have a
TP-SAGB \citep[e.g.,][]{Tauris2015b}.}  We develop a simple, \textit{ad
  hoc} procedure that halts the models before the onset of thermal
pulses or dredge-out.

We stop the evolution and eject the envelope once two conditions are
met simultaneously.  First, we check if the highest temperature in the
region of the star where the energy generation rate is dominated by
CNO-cycle hydrogen burning is greater than roughly
$\unit[3 \times 10^7]{K}$.  This condition is
naturally triggered as the hydrogen-burning shell develops before the first thermal pulse and is
necessarily triggered
before high-temperature proton ingestion can lead to numerical
difficulties.  Second, we check that the CO core mass and the ONe core mass
are within 10\% of each other.  This condition ensures that we do not
stop the evolution before the ONe core has finished forming.  At this
point, we remove the envelope using an artificially-enhanced stellar
wind with $\dot{M} \sim \unit[10^{-2}]{\Msunyr}$.  We turn off nuclear
burning during this phase for numerical simplicity.  After the
envelope has been removed we are left with a hot ONe WD model with a
thin hydrogen and helium envelope.

Our goal is to produce a set of WD models that have a range of
qualitatively different chemical profiles.  Therefore, like
\cite{Farmer2015ApJ}, we vary a handful of stellar parameters
including initial mass, rotation at the zero-age main sequence (ZAMS),
and the location and strength of convective overshooting.

We run a set of non-rotating models without any
convective overshoot (series NRNO).  We run a set of models (series
RO) with the fiducial rotation and overshooting parameters adopted by
\citet{Farmer2015ApJ}, which includes overshoot at all convective boundaries. The propagation of the carbon flame, and therefore the
subsequent chemical profile, is sensitive to the treatment of
mixing at the convective boundary near the flame  \citep{Siess2009, Denissenkov2013b}.  To explore this effect, we also run a set of rotating models (series RMO) with an overshooting treatment where mixing below convective regions associated with carbon burning does not occur.  This corresponds to the \MESA\ options
\begin{verbatim}
	overshoot_f_below_burn_z_shell  = 0.000
	overshoot_f0_below_burn_z_shell = 0.000
\end{verbatim}
This choice is motivated by the
findings of \cite{Lecoanet2016ApJ} who performed idealized hydrodynamic simulations of a carbon flame and found that little mixing occurs at this interface as convective plumes were not
able to penetrate into the carbon flame.

\subsection{Results of SAGB \MESA\ models}
\label{sec:mesa-agb-results}

\begin{deluxetable*}{ cc|ccc|ccc  }

  \tablecaption{Summary of ONe WD models. Models in a series are evolved with an identical treatment of rotation and convective overshooting. The left part of the table shows the model identifiers.  The middle part shows initial conditions and model parameters.  The right part shows key properties of the formed WDs. Note that the WD progenitor models had their evolution artificially truncated before the TP-SAGB. If they were allowed to evolve though the TP-SAGB, the mass of the WD might be somewhat different (depending on the efficiency of third dredge-up). \label{table:stellar_models}}

	\tablehead{
          Series & Model & Initial Mass  & Rotation  & Overshooting  & WD Model Mass & \multicolumn2c{Central $^{12}$C [\%]} \\ & & [$M_\odot$] & [$\omega/\omega_{\rm crit}$] & ($f_{\rm ov}$) & Mass [$M_\odot$] & pre-cool & post-cool }
        
	\startdata
	RMO & 1     & 7.2  & 0.25 & \enspace 0.016 / 0.000 & 1.06 & 6.99 & 2.69 \\ 
	    & 2     & 7.6  &      &       & 1.18 &  0.98  &  0.69  \\
	    & 3     & 8.0  &      &       & 1.28 &  0.22  &  0.18  \\ 
	    & 4     & 8.4  &      &       & 1.34 &  0.25  &  0.07  \\ \hline
	RO  & 5     & 7.2  & 0.25 & 0.016 & 1.08 &  32.72 &  16.30 \\ 
	    & 6     & 7.6  &      &       & 1.17 &  15.76 &  2.89  \\
	    & 7     & 8.0  &      &       & 1.28 &  3.94  &  1.02  \\ 
	    & 8     & 8.4  &      &       & 1.34 &  0.31  &  0.09  \\ \hline
	NRNO& 9     & 9.5  & 0.00 & 0.000 & 1.01 &  2.46  &  2.65  \\ 
	    & 10    & 10.0 &      &       & 1.09 &  4.23  &  1.64  \\
	    & 11    & 10.5 &      &       & 1.14 &  2.16  &  1.19  \\ 
	    & 12    & 11.0 &      &       & 1.19 &  0.68  &  0.63  \\
	\enddata

\end{deluxetable*}

\begin{figure*}
	\centering
	\includegraphics[width=\textwidth]{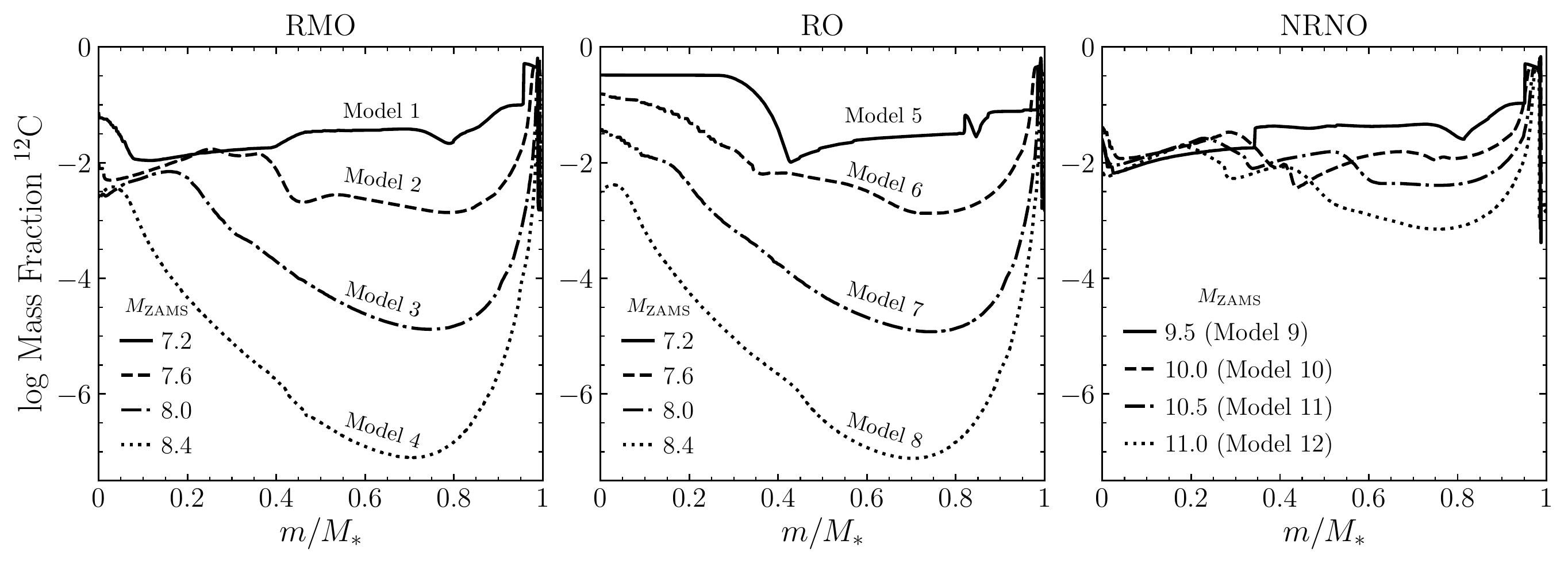}
	\caption{Carbon profiles as a function of normalized mass coordinate for all 12 WD models (before cooling).
 Each panel shows a group of WD models from stars of different initial (ZAMS) masses that use identical stellar modeling parameters.
          The initial conditions and parameters for each model are indicated in Table \ref{table:stellar_models}.}
	\label{fig:comparing_carbon_profiles}
\end{figure*}


We produced a total of 12 WD models.  These models and their identifiers are listed in Table \ref{table:stellar_models} together with the key parameters including initial mass, rotation at ZAMS, and convective overshooting. All but one model produce massive ONe WDs. Model 5 produces a hybrid CO-ONe WD.

\begin{figure}
	\centering
	\includegraphics[width=\columnwidth]{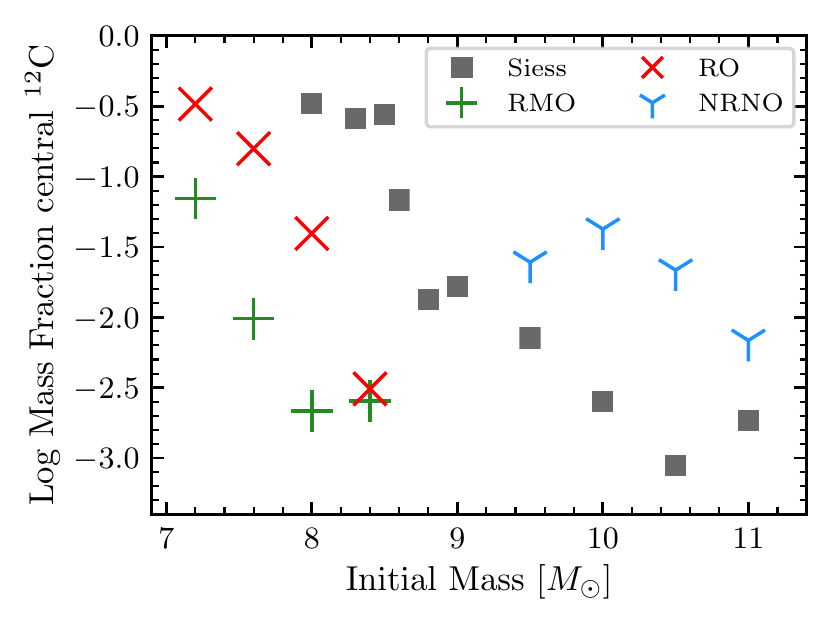}
	\caption{Central carbon abundance in SAGB models as a function of initial mass.  Our \MESA\ models (colored symbols) are split into three different groups with differing stellar controls including initial mass, rotation, and overshooting (see Table \ref{table:stellar_models}).
For comparison, the SAGB models of \citet{Siess2007A&A} are shown as the gray squares.}
\label{fig:comparing_central_carbon}
\end{figure}

Convective overshooting and rotation can have large effects on the chemical profile of the WD and especially that of $^{12}$C.
Figure~\ref{fig:comparing_carbon_profiles} shows the carbon profiles in our models; each panel compares different initial masses within the same model series.  Broadly, models from more massive stars have less carbon than those from lower mass stars.\footnote{This is why \carbon[12] is generally not important in single-star models for ECSNe which are necessarily more massive than those that produce ONe WDs.  For example, after the end of core carbon burning, the ECSN progenitor model of \citet{Takahashi2013} has a \carbon[12] mass fraction $\approx 10^{-3}$.}
This trend is set as models transition from flames that fail to reach the center (leaving lots of unburned carbon) at lower mass
  to central carbon ignitions (with almost complete carbon burning) at higher mass.
The detailed behavior in between these extremes is complex, with a mixture of carbon flames and flashes \citep[see e.g., Figure 3 in][]{Farmer2015ApJ}.
This trend is particularly pronounced in the models that include overshooting (left and center panels).
Models 1-8 from series RMO and RO have carbon depletion especially at larger radii while models 9-12 from series NRNO have comparatively homogeneous carbon profiles. 

Stellar models that exhibit efficient convective overshooting below the carbon flame tend to extinguish the carbon flames earlier resulting in an increased abundance of carbon in the core.  This effect is clearly visible in the left and center panels of Figure \ref{fig:comparing_carbon_profiles}.  The profiles of series RO and series RMO, which differ only by their assumption about mixing below the carbon flame, have similar carbon profiles in the outer half of the cores.  However, series RO has systematically higher carbon abundances in the inner regions than series RMO.  We find that all 12 WD models exhibit qualitative differences in their carbon profiles depending on their initial masses and modeling assumptions.

\subsection{Comparison to Siess Models}
\label{sec:siess-comparison}

\cite{Siess2007A&A} constructed SAGB models using the stellar
evolution code STAREVOL.  These models do not take into account the
effects of rotation.  Like our \MESA\ models, they use a decaying
exponential treatment for convective overshooting with the same value
of \mbox{$f_{\rm ov} = 0.016$}, though this is only applied at the
edge of the convective core.  We compare our models with their solar
metallicity ($Z = 0.02$) models that include convective overshooting.

Figure~\ref{fig:comparing_central_carbon} shows the central $^{12}$C abundance at the end of carbon burning as a function of initial mass. The models from \cite{Siess2007A&A} are shown as gray squares. Models with higher initial masses have lower central carbon fractions. Our \MESA\ models exhibit similar trends. Our models are not directly comparable to the Siess models at fixed initial mass because the different assumptions about rotation and overshooting imply a different mapping between initial mass and core mass.
However, these independent sets of SAGB models display the same feature that stars in this transition mass range show a broad range of central carbon mass fractions ranging from tenths of a percent to tens of percent.

Figure~\ref{fig:comparing_central_carbon} also illustrates the
sensitivity of the central carbon fraction to convective overshooting
below regions associated with carbon burning.  The series RO models
have central carbon mass fractions up to a factor of 10 greater than
those in series RMO at fixed initial mass; again, the only difference
between these models is the overshooting parameter at the lower boundary of the convection zone immediately above the carbon flame.

\section{Homogenization during WD Cooling}
\label{sec:homogenization}

In order for the ONe WD to evolve towards collapse, it must eventually
accrete from a companion. This interaction is unlikely to begin
immediately after the formation of the WD, but must instead wait for
the companion star to evolve and fill its Roche lobe.  Therefore, our
modeling assumes that after formation there is a phase of evolution in
which the WD does not interact with its companion and simply
cools as an effectively single object.

As a WD cools and approaches an isothermal configuration, mixing can
smooth out compositional gradients in the WD interior.\footnote{
While we describe these mixing processes as
  occurring during WD cooling, in models that self-consistently follow the evolution
  through the TP-SAGB, they may have already begun to act during that
  phase.}
Often in the
literature this process is referred to as ``rehomogenization''---framed as Rayleigh-Taylor instabilities
smoothing regions with negative molecular weight gradients---and is
often assumed to occur quasi-instantaneously \citep[e.g.,][]{Althaus2007,
  Camisassa2018}.  However, more formally, the presence of a destabilizing
composition gradient leads to mixing via thermocompositional
convection. Depending on the magnitude of the stabilizing
temperature gradient, this can take the form of overturning convection or
double-diffusive fingering (thermohaline) convection.  This distinction is important
as these different processes occur over different timescales.

In CO WDs, the evolutionary effects of this kind of mixing are generally minor.
However in hybrid CO-ONe WDs, where an ONe mantle overlays a CO core, such mixing is of critical importance
as it will destroy the stratified core-mantle structure
on a timescale significantly shorter than the likely timescale for the WD to grow to
the Chandrasekhar mass \citep{Brooks2017ApJ}.  The complex carbon
profiles at formation in our ONe WD models (see Section~\ref{sec:mesa-agb-results}) will similarly be affected by these
mixing processes and this can substantially alter the central carbon
fraction.

\begin{figure}
	\centering
	\includegraphics[width=\columnwidth]{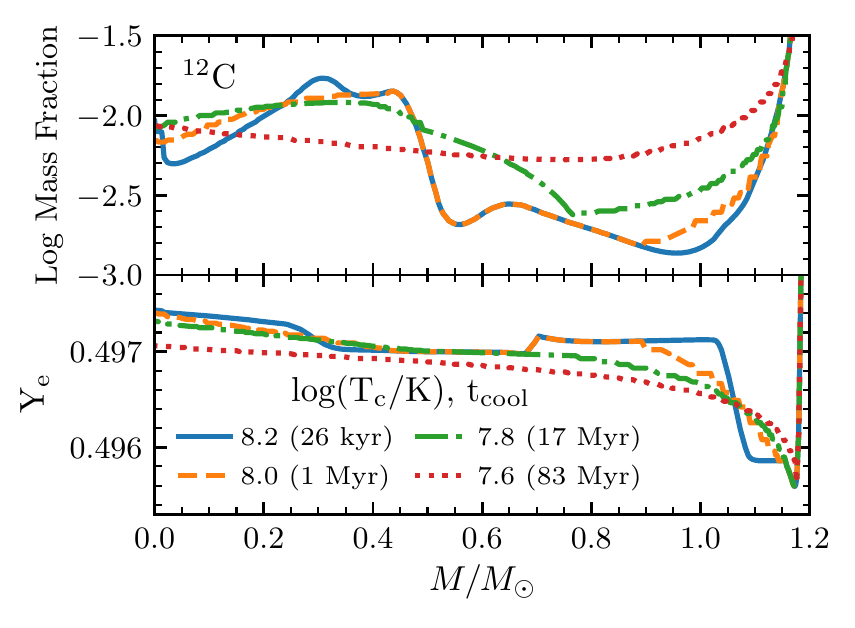}
	\caption{Evolution of the chemical profiles during WD cooling
          {\it without} thermohaline mixing.  The upper panel shows
          the $^{12}$C mass fraction and the lower panel shows \Ye,
          both as a function of the Lagrangian mass coordinate.
          Profiles are shown at a sequence of central temperatures; the
          time to reach the given temperature is indicted in
          parenthesis.}
	\label{fig:profile_ye_noThermo}
\end{figure}

\begin{figure}
	\centering
	\includegraphics[width=\columnwidth]{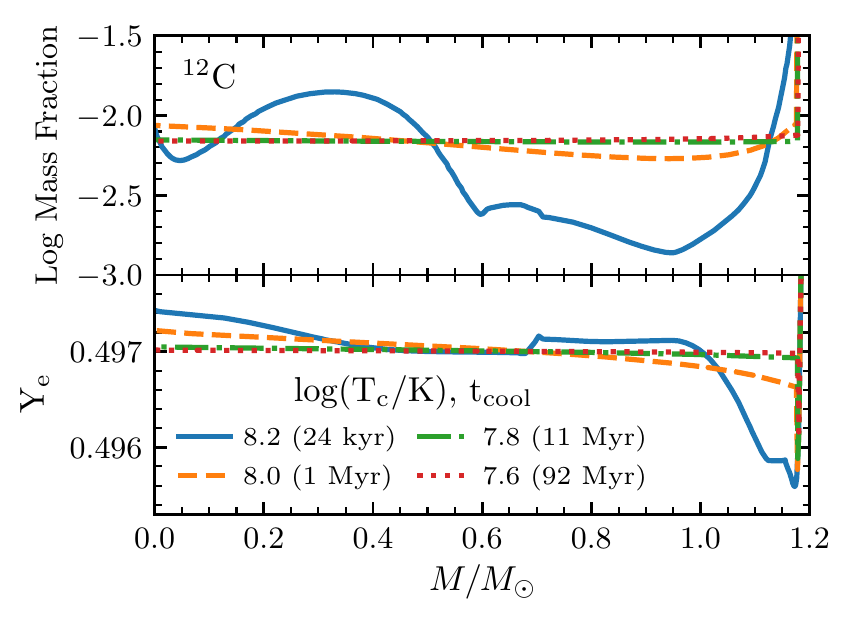}
	\caption{Evolution of the chemical profiles during WD cooling
          {\it with} thermohaline mixing (cf.~Figure~\ref{fig:profile_ye_noThermo}).}
	\label{fig:profile_ye_Thermo}
\end{figure}

In order to illustrate the effects of the mixing processes, we show the evolution of our WD model 2 (see Table~\ref{table:stellar_models}) during cooling.  Figure \ref{fig:profile_ye_noThermo} shows the model evolving assuming no thermohaline mixing occurs while Figure \ref{fig:profile_ye_Thermo} shows the model evolving including thermohaline mixing.  The lower panels of these plots show the electron fraction (\Ye): as the temperature becomes isothermal, convection alone will leave behind the neutrally buoyant \Ye\ profile while the addition of thermohaline mixing leads to a flat \Ye\ profile.  The upper panels show the effects that mixing has on the $^{12}$C abundance.  Figure \ref{fig:profile_ye_noThermo} shows that without thermohaline mixing, a significant gradient in the $^{12}$C profile and \Ye\ persists after 80 Myr.  In contrast, the WD model with thermohaline mixing
shown in Figure \ref{fig:profile_ye_Thermo} has already reached a similar level of homogenization after 1 Myr and has almost completely homogeneous $^{12}$C and \Ye\ profiles after only 11 Myr.  This behavior is similar to what was found in \cite{Brooks2017ApJ}.

Moving forward, all of our WD models are evolved using thermohaline mixing, as this is an expected physical effect.  We do not include the effects of phase separation at crystallization which may further affect the central abundances \citep{Segretain1993}; such capabilities are not presently included in \MESA.  Since \carbon[12] has the lowest charge of the isotopes present, the effect of phase separation would likely be to reduce the central carbon abundance.

\begin{figure}
	\centering
	\includegraphics[width=\columnwidth]{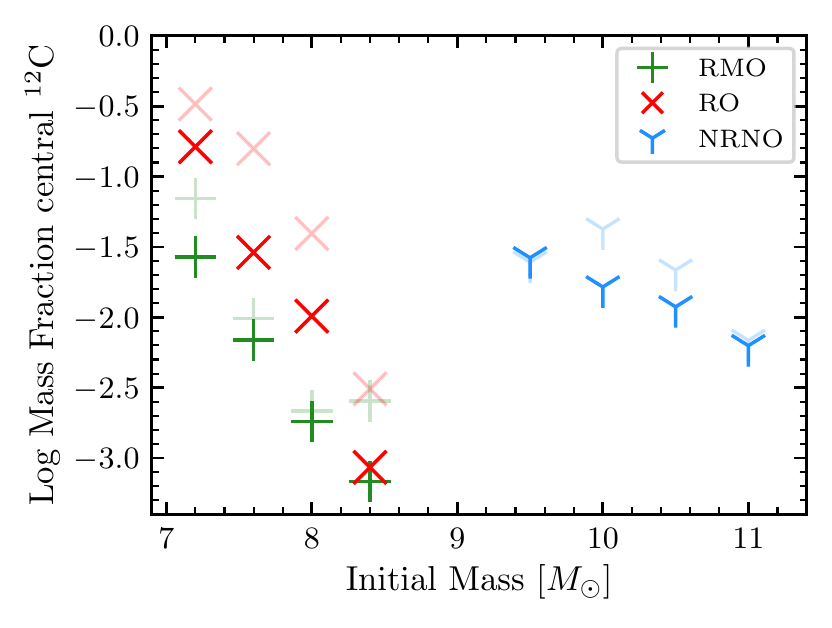}
	\caption{Central carbon abundance as a function of the initial stellar mass after the WD models have cooled for $\approx \unit[100]{Myr}$. Translucent points indicate central carbon abundances prior to cooling (same as Figure~\ref{fig:comparing_central_carbon}).}
	\label{fig:comparing_central_carbon_post_cool}
\end{figure}

After $\approx 100$ Myrs of cooling, all our WD models exhibit effectively homogeneous interior chemical profiles.  Figure~\ref{fig:comparing_central_carbon_post_cool} shows the central mass fractions at this time; this value now reflects the abundance throughout the interior.  The pale symbols show the mass fractions before cooling, illustrating that the central abundance of isotopes like $^{12}$C can change by a factor of a few. (The values pre- and post-cooling are also reported in Table~\ref{table:stellar_models}.)  In Section~\ref{sec:how-much-C}, we showed that varying initial mass, convective overshoot, and rotation produced ONe WDs having chemical profiles with qualitatively different shapes, but here we have shown that those differences are subsequently erased.  However, Figure~\ref{fig:comparing_central_carbon_post_cool} shows that the overall range of central carbon abundance across models persists.

\section{Effects of Residual Carbon}
\label{sec:carbon}

In Sections~\ref{sec:how-much-C} and \ref{sec:homogenization},
we showed that ONe WD models from SAGB star evolution
using \MESA\ had residual \carbon[12] mass fractions $X_{12} \approx 10^{-3}- 10^{-1}$.
In this section, we will outline the role of \carbon[12] in the evolution of the WD towards collapse.  
Similarly, \citet{Gutierrez2005} evolve a set of ONe WD models with parameterized
\carbon[12] mass fractions in the range $X_{12} = 0.01 - 0.06$,
motivated by the unburnt carbon in SAGB models
\citep{Dominguez1993, Ritossa1996, GilPons2001, GilPons2002}.
They find that that models with relatively small \carbon[12] abundances
$(X_{12} \gtrsim 0.015)$ can allow for ignition at densities
$\sim \unit[10^9]{\gcc}$. Ignition at these low densities seems likely lead to disruption of the
star.  Indeed, if carbon burning occurs before the exothermic electron
captures, the evolution will be more analogous to the simmering phase
experienced by near-Chandrasekhar mass CO WDs in which carbon burning
leads to the development and growth of a large central convection zone
\citep[e.g.,][]{Lesaffre2006, Piro2008b}.  However, as we will
demonstrate shortly, the presence of Urca-process cooling due to \sodium[23] makes it
unlikely that carbon ignition occurs in advance of the exothermic
electron captures on \magnesium[24] and \sodium[24].

\subsection{Carbon Ignition and Electron Captures}
\label{sec:crit-curves}

The models in \citet{Gutierrez2005} start with central conditions
$\rhoc = \unit[10^9]{\gcc}$ and $\Tc = \unit[2\times 10^8]{K}$.  From
their Figure 3, it appears that these models almost immediately begin
carbon burning.  For lower carbon mass fractions, the carbon-burning
runaway takes slightly longer, allowing additional compression to
occur during the temperature increase.  The lowest carbon mass
fraction model ($X_{12} = 0.01$) appears to run out of fuel before
reaching temperatures $\gtrsim \unit[10^9]{K}$ and then subsequently
rejoins the compression-driven evolution towards higher density and the eventual \neon[20] electron
captures that is characteristic of their carbon-free models.

The fact that the models of \citet{Gutierrez2005} appear to be burning
carbon at or near their initial central conditions is puzzling.
Figure 4 of \citet{Gasques2005} shows ignition lines (defined
as where the energy generation rate is equal to the thermal neutrino
losses) and burning
timescales for pure carbon that cover the relevant range of densities
and temperatures.  The initial central conditions from
\citet{Gutierrez2005} are well below even a burning timescale of order
a Hubble time.

\begin{figure}
  \centering
  \includegraphics[width=\columnwidth]{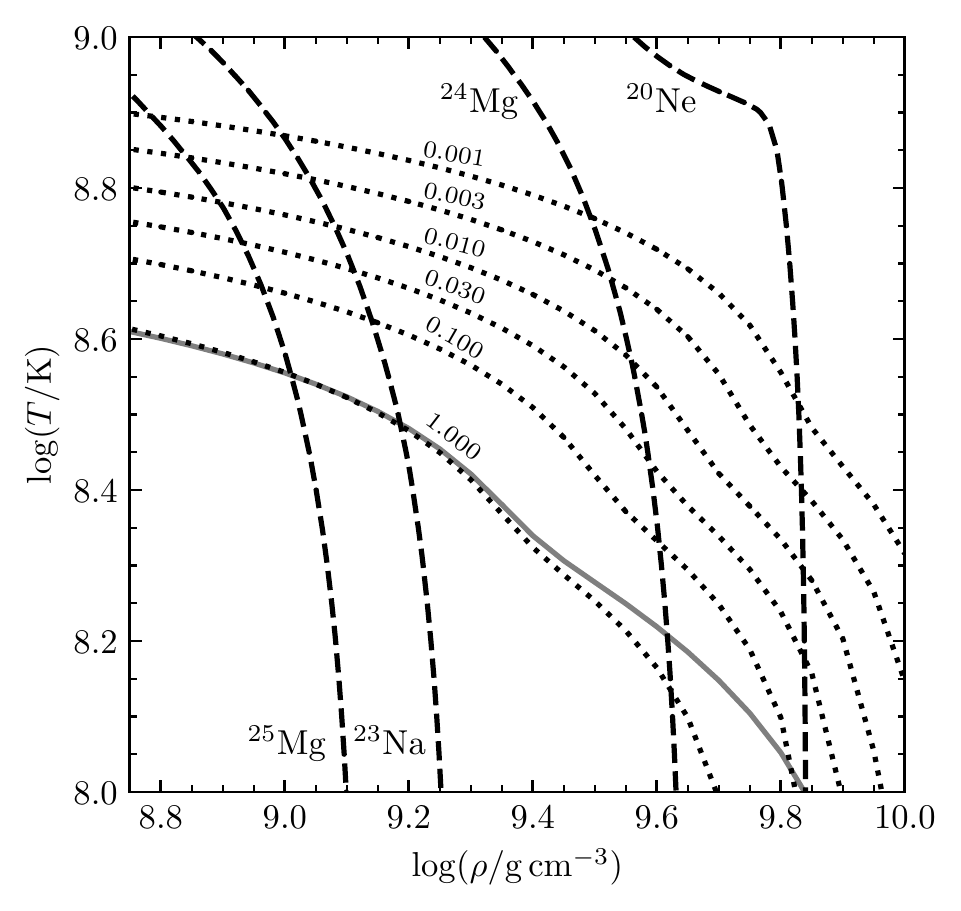}
  \caption{Critical lines for carbon burning and electron captures.
    The dotted lines show the carbon ignition curves for the
    input physics adopted in our \MESA\ models, labeled by
    carbon mass fraction.  The solid gray line shows the ignition
    curve for pure carbon using the rate from \citet{Gasques2005}.
    The dashed lines show where the indicated isotope has an electron
    capture timescale of $\unit[10^4]{yr}$.}
  \label{fig:timescales}
\end{figure}

Figure~\ref{fig:timescales} shows the carbon ignition lines  for a range of carbon mass fractions.  These are evaluated
with the same input physics as our subsequent \MESA\ models. We use
the carbon burning rate from \citet{Caughlan1988} and the
``extended'' screening treatment in \MESA\ as described in
\citet{Paxton2011}.  This combines the results of \citet{Graboske1973} in the weak
regime and \citet{Alastuey1978} with plasma parameters from
\citet{Itoh1979} in the strong regime.  The neutrino loss rates are
calculated via the fitting formulae of \citet{Itoh1996a}.  For
comparison, the solid grey line shows the \MESA\ implementation of the
rate given by \citet{Gasques2005} for pure carbon\footnote{The extension of these results to multicomponent plasmas and hence
smaller carbon mass fractions is given by \citet{Yakovlev2006}, but this is not presently implemented in \MESA.}
($X_{12} = 1.0$).
At low temperature (roughly in the regime where $T$ is below the ion
plasma temperature), these ignition curves disagree, reflecting
differences in the screening treatment.  Figure 4 of
\citet{Gasques2005} shows the theoretical uncertainty associated with
the ignition curves in this regime; the pure carbon \MESA\ ignition
curve in Figure~\ref{fig:timescales} lies within that range.

Figure~\ref{fig:timescales} also indicates where the key weak
reactions will become important.  The dashed lines show where the
electron capture timescale for each of \magnesium[25], \sodium[23],
\magnesium[24], and \neon[20] is equal to a characteristic compression
timescale of \unit[$10^4$]{yr} \citep{Schwab2015}, with more rapid
electron captures occurring to the right of the line.  These curves
were evaluated using the rate tabulations of \citet{Suzuki2016} assuming
$\Ye = 0.5$.

\subsection{Simple MESA Models}

In order to illustrate the potential role of carbon, we follow the
approach of \citet{Schwab2015, Schwab2017a} by constructing
quasi-homogeneous WD models with parameterized abundances.  This
provides a simple way to isolate the effects of \carbon[12].  In
Section~\ref{sec:to-collapse} we will apply this understanding to the
more realistic WD models described in previous sections.  The
calculations shown in this section use \MESA\ version 10108.  All
input files will be made publicly available at
\url{http://mesastar.org}.

\begin{table}
  \centering
  \caption{The set of compositions used in our \MESA\ models.  Each
    composition is referenced in the text by the identifier listed in
    the top row.  Each column lists the mass fractions of the isotopes
    that were included.  Dashes indicate that a
    particular isotope was not included.  A range of \carbon[12] mass
    fractions were considered; the \neon[20] mass fraction was chosen
    to ensure the mass fractions sum to unity.}
  \label{tab:compositions}
  \begin{tabular}{r|cc}
    \hline
    Isotope & SQB15C & SBQ17C  \\
    \hline
    \carbon[12]    & 0.0 - 0.1 & 0.0 - 0.1  \\
    \oxygen[16]    & 0.50 & 0.50  \\
    \neon[20]      & remainder & remainder  \\
    \sodium[23]    & ---   & 0.05  \\
    \magnesium[24] & 0.05 & 0.05  \\
    \magnesium[25] & ---   & 0.01  \\
    \hline
  \end{tabular}
\end{table}

We first create a pre-main sequence star of $\unit[1.0]{\Msun}$
with a solar composition.
We select this as a starting point because a high-entropy model is
more amenable to our relaxation procedure than a highly degenerate WD
model.  During these relaxation phases, we disable nuclear reactions.
We evolve until the model reaches a central density of $\logRhoc = 4$
and then
we relax its composition to one of the compositions indicated in
Table~\ref{tab:compositions}.  These are the fiducial compositions
from \citet{Schwab2015, Schwab2017a} plus a variable amount of carbon.
We continue the evolution, with the model cooling and contracting,
until it reaches a central density of $\logRhoc = 7$.
We then have the models accrete a 50/50 oxygen-neon mixture%
\footnote{We accrete material that is free of hydrogen, helium and carbon in order to avoid the complications associated with following the nuclear burning on the surface of the WD.}
at a rate
$\unit[10^{-6}]{\Msunyr}$ until the WD has reached $\logRhoc = 8.6$.
This value is selected to be below the density at which
electron captures on the isotopes present in the initial composition will first occur.
This procedure creates near-Chandrasekhar mass WD models with
homogeneous inner regions of the desired compositions.

We then take these models and continue letting them accrete at
$\unit[10^{-6}]{\Msunyr}$.  This choice represents an accretion rate
that is roughly characteristic of any near-Chandrasekhar WD accretor
stably burning hydrogen or helium \citep[e.g.,][]{Wolf2013, Brooks2016}.
The long period of accretion in these models means that the
  memory of their initial central temperatures are erased as the
  models come into a balance where the compressional heating balances
  neutrino losses \citep{Paczynski1971d, Schwab2015, Brooks2016}.
  However, this does imply that the temperature of these models is set
  by our choice of accretion rate.
We evolve until the model experiences
carbon ignition (defined as the energy release from carbon burning
locally exceeding thermal neutrino losses).
We note that some models, in particular those with carbon
  fractions $\la 0.01$, may evolve beyond the carbon ignition lines
  corresponding to their initial carbon abundances. This is because by
  the time an evolutionary model reaches the temperature and density
  of the ignition curve corresponding to its initial carbon abundance,
  some of the carbon can have already been consumed.

\begin{figure}
  \centering
  \includegraphics[width=\columnwidth]{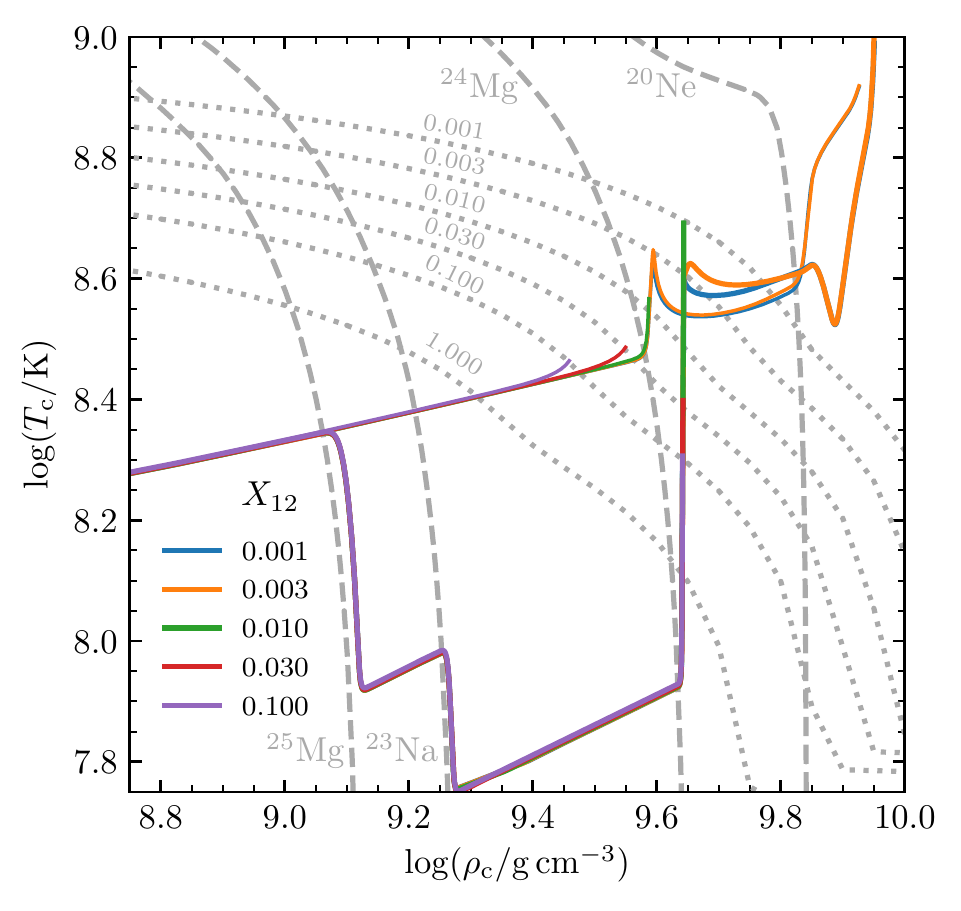}
  \caption{Evolution of simplified WD models with a range of carbon
    fractions.  The thin lines show models without Urca cooling
    (composition SQB15C); the thick lines show models with
    Urca-process cooling (composition SBQ17C). The light grey lines in
    the background are the same as those in
    Figure~\ref{fig:timescales}.
    Models are halted when the energy release from carbon burning
      locally exceeds thermal neutrino losses.}
  \label{fig:toy-models}
\end{figure}

\citet{Schwab2017a} demonstrated that models that experience electron
captures after significant Urca-process cooling has occurred can
develop convectively-unstable regions near the electron capture
front.  Proper modeling of this convection likely requires both
theoretical and numerical progress and remains an important open
question.  We discuss the difficulties of evolving \MESA\ models in
this phase more specifically in Section~\ref{sec:to-collapse}.  To circumvent these
issues, we artificially suppress the action of convection in the
\MESA\ models using the control  \texttt{ mlt\_option = 'none'}.
This is an important caveat that applies to the evolutionary tracks shown in Figure~\ref{fig:toy-models} after \magnesium[24] electron captures $(\logRho \ga 9.6)$.  Nonetheless, these models still provide insight into
when carbon burning first begins to affect the evolution.

Figure~\ref{fig:toy-models} shows the evolution of the central temperature and density in the models.  The
thin lines show the results of models with the SQB15C composition, which do not
include Urca-process cooling since there is no \sodium[23] or \magnesium[25].  These models are less physically
realistic, but we show them primarily for comparison with the
results of \citet{Gutierrez2005}.
Models with $X_{12} \ga 0.02$
experience carbon ignition at densities below where electron captures on \magnesium[24] would occur.
Energetically, it requires  burning
approximately $\unit[0.02]{\Msun}$ of
carbon to heat the WD to the conditions for dynamical burning
\citep{Piro2008a}.  Therefore, since these models have roughly that much carbon, if we continued their evolution they may be able to reach the conditions for thermonuclear explosions.
Note that the $X_{12} = 0.1$ model is similar to what one would expect from a fully-mixed hybrid CO-ONe WD.
Models with very low carbon
abundances $X_{12} \la 0.003$ evolve much like carbon-free models;
the thin orange and blue lines are nearly identical and continue
until oxygen ignition.
Intermediate abundance models with $0.003 \la X_{12} \la 0.02$
experience carbon ignition as a result of the $A=24$ electron
captures.  This range encompasses the carbon fractions in many of the
ONe WD models shown in Section~\ref{sec:homogenization}.

However, the presence of Urca-process cooling changes this picture,
causing models across a range of compositions and accretion rates to
reach $\Tc \la \unit[10^8]{K}$ \citep[see Equation 15
in][]{Schwab2017a}. The thick lines in Figure~\ref{fig:toy-models}
show the results of models with composition SBQ17C; the presence of
\sodium[23] and \magnesium[25] leads to significant cooling.  Models
that experience this cooling will therefore almost certainly reach the
\magnesium[24] electron captures before igniting carbon.  Models with
very low carbon abundances $X_{12} \la 0.003$ evolve much like
carbon-free models (though again, we note that these models do develop
convectively unstable regions that are artificially not allowed to
mix).  Models with $X_{12} \ga 0.01$ experience carbon ignition as a
direct result of the $A=24$ electron captures.

Stellar models indicate significant abundances of the odd mass number
isotopes (in particular \sodium[23]) that are responsible for the
Urca-process cooling.  Thus the SBQ17C models are the ones that should
guide our understanding of WDs with self-consistently set
compositions.  Urca-process cooling prevents carbon
ignitions from occurring in advance of the exothermic electron captures on
\magnesium[24].  The
SBQ17C models do show carbon ignition triggered by the $A=24$ electron
captures above a critical carbon fraction of around 1 per cent by
mass.  Since this is within the range seen in our self-consistent
stellar models of ONe WDs, this is an indicator that carbon may play
an important role in at least some systems.


\section{Evolution to Collapse}
\label{sec:to-collapse}

In this section, we continue the schematic binary evolution of the WD
models produced in Sections~\ref{sec:how-much-C} and
\ref{sec:homogenization} and examine the role of carbon in the
evolution towards their final fates.

\subsection{Setup of accreting WD models}

In the overall evolutionary scenario for AIC, the ONe WD will stop cooling once it begins accreting from its companion via Roche lobe overflow. This can be achieved through expansion of the secondary driven by stellar evolution processes or a shrinking of the orbit through angular momentum losses.  Either process results in a massive WD accreting material.  

We model the effects of accretion from a companion by adding a
constant accretion rate of \oxygen.  We again select a rate of
$\unit[10^{-6}]{\Msunyr}$ as roughly characteristic of any
near-Chandrasekhar WD accretor stably burning hydrogen or helium.
Accreting oxygen simplifies the models as it avoids having to follow
the numerically challenging hydrogen, helium, or carbon flashes that
could occur during the accretion process.  We turn rotation off on all
models to avoid numerical problems with angular momentum transport in
the WD.  This is consistent with past work that has primarily studied non-rotating WDs.

We use a version of the original nuclear network 
  \texttt{sagb\_NeNa\_MgAl.net} modified to add the isotopes \oxygen[19],
\oxygen[20], \fluorine[20], \fluorine[21], \fluorine[23], \neon[23],
\neon[24], \sodium[24], \neon[25], \sodium[27], and \magnesium[27] and
the weak reactions that link them to the isotopes originally present in
our network.  We use the reaction rate tables from \cite{Suzuki2016}
which are specifically calculated for the conditions in high-density
ONe cores.  We adopt EOS options\footnote{ \citet{Schwab2017a} include
  all isotopes in the nuclear network in the PC \citep{Potekhin2000}
  EOS calculations by using the control
  \texttt{mass\_fraction\_limit\_for\_PC = 0d0}.  In \MESA\ r9793
  doing so with networks including neutrons causes the program to halt
  unexpectedly.  We backport the one-line fix from \MESA\ commit
  r10361 to our version.}  and spatial/temporal resolution controls similar to those
of \cite{Schwab2017a}, as it is demonstrated that those choices lead
to numerically converged models (see their Appendix B).

\subsection{Model evolution}

\begin{figure}
  \centering
  \includegraphics[width=\columnwidth]{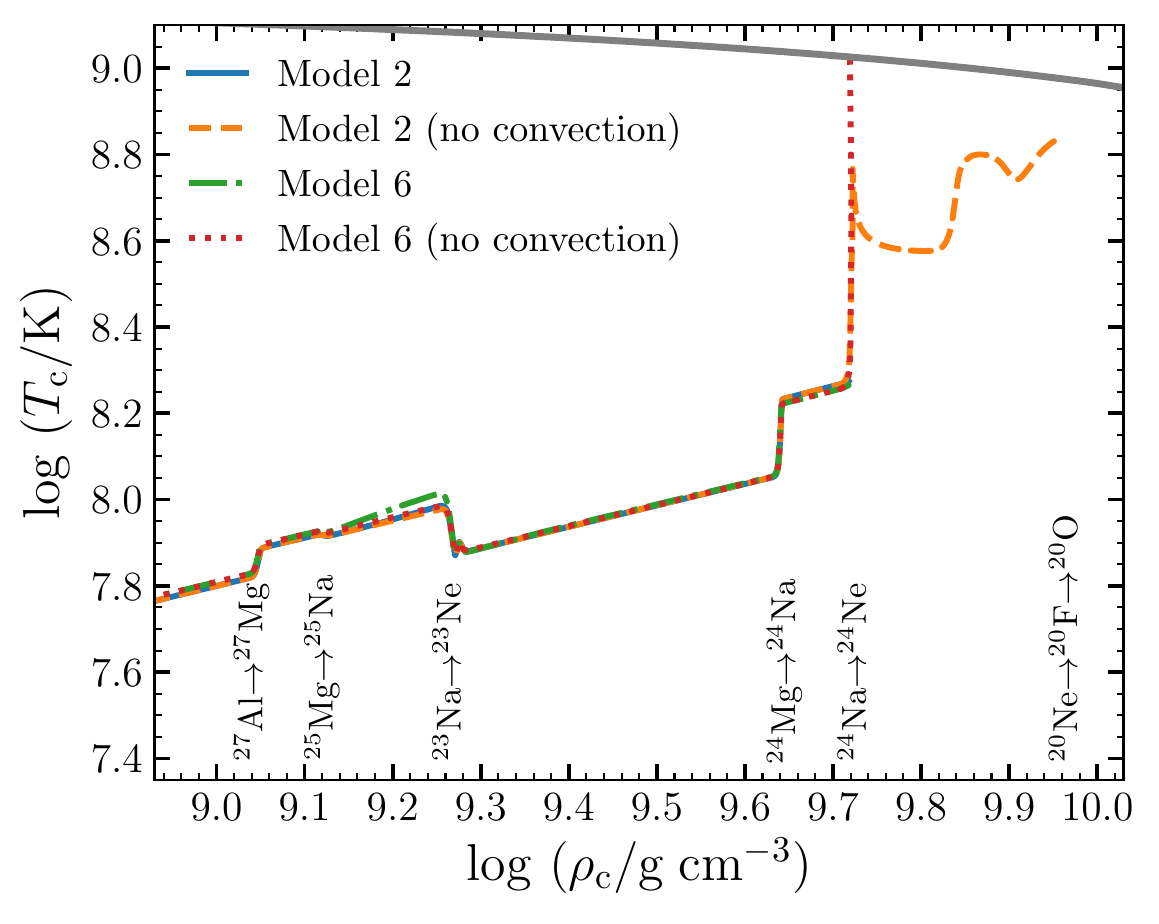}
  \caption{Central evolution of a lower carbon fraction model (model
    2, $X_{12} \approx 0.007$) and a higher carbon fraction model
    (model 6, $X_{12} \approx 0.03$).  Annotations indicate the
    densities at which key weak reactions occur.  We show models
    evolved normally (which terminate at the onset of electron
    captures on \sodium[24]) and models evolved without convection
    (which terminate roughly at oxygen ignition).  The gray line is
    the oxygen ignition curve.}

  \label{fig:rhoT-WD}
\end{figure}

We show the evolution of the central density and temperature of a
subset of our models in Figure~\ref{fig:rhoT-WD}.  In particular, we
select models 2 and 6 from Table~\ref{table:stellar_models}.  These
models both descend from \unit[7.6]{\Msun} ZAMS models, with the only
difference between them being the efficiency of overshooting
associated with the carbon flame.  These produce similar WD models
that differ mainly in their carbon abundance; model 2 and model 6 have central
carbon mass fractions of 0.007 and 0.03 respectively.

Compared to the toy models in Section~\ref{sec:carbon}, these models
begin somewhat cooler.  This is because the toy models had been
steadily accreting since they were much lower mass and had reached a
balance between compressional heating and neutrino cooling
\citep{Paczynski1971d}.  In contrast, our realistic ONe WD models
had already cooled below this temperature, and being 
initially more massive, have not accreted for long enough to reheat.

The annotations on the lower part of the figure indicate the location
of key weak reactions.  We see some difference from the toy models
shown in Figure~\ref{fig:toy-models}.  Mildly exothermic captures on
\aluminum[27] occur; this species is not included in our toy models.
The Urca-process cooling due to \magnesium[25] is less than in the toy
models, due both to the initially lower temperature and the lower
abundance of this isotope.  The cooling from \sodium[23] is more
dramatic, though the small glitch near the end of cooling is an artifact
of the reaction rate tables and does correspond to slightly too little
cooling \citep[see Appendix D in][]{Schwab2017a}.  The separation (in
density) of the \magnesium[24] and \sodium[24] electron captures is
caused by different assumed strengths of the (currently unmeasured)
non-unique second forbidden transitions \citep[see Section 5
in][]{Schwab2017a}.

All model tracks are similar up until the electron captures on
\sodium[24].  The heating from these is sufficient to ignite carbon.
We are unable to evolve the models including the effects of convection
beyond this point.  By artificially suppressing convection, we are
able to continue their evolution.  These tracks are marked ``no convection'' in Figure~\ref{fig:rhoT-WD}. Subsequently, the difference
between the two model tracks is apparent, with the a bifurcation
reflecting the difference in the carbon abundance.  The low carbon
fraction model 2 continues to high density while the low carbon
fraction model 6 experiences a relatively low density carbon-assisted
oxygen ignition.

\subsection{Convective Instability and Carbon Burning}

Both in Section~\ref{sec:carbon} and in this section, we encountered
numerical difficulties when convective regions developed in the
regions of the star undergoing electron-capture reactions.  We
circumvented this problem by artificially turning off convection.  The
further evolution of the models is thus unphysical, though hopefully
remains schematically useful.

To understand why it is difficult to evolve the models beyond this
point, it is first important to recall why these models become
convective.  The onset of exothermic electron captures leads to a
strongly destabilizing temperature gradient.  However, this is locally
accompanied by the development of a strongly stabilizing composition
(electron fraction) gradient.  When using the Ledoux criterion, this
can lead to overall convective stability \citep[e.g.,][]{Miyaji1987,
  Schwab2015}.  The crossed temperature and composition gradients mean
that this region is unstable to double diffusive convection, though
this process is thought to operate too slowly to have a significant
effect given the short evolutionary timescale.  When including the
effects of Urca-process cooling, the models of \citet{Schwab2017a}
became convectively unstable because thermal conduction causes a more
rapid spread of the thermal gradient and leads to a situation in
which the thermal and compositional gradients can no longer balance
locally because their spatial extents are no longer well-matched.  We
emphasize that the development of these convective regions is
unrelated to the presence of \carbon[12] and occurs even in
carbon-free models.

However, once the exothermic electron captures raise the temperature
enough for carbon burning to begin, its energy release can further
contribute to convective instability.\footnote{The heating from carbon
  burning is accompanied by some stabilizing change in composition, as
  immediate electron captures on carbon-burning products can occur at
  these densities.  Since these electron captures are super-threshold
  they also deposit additional thermal energy.  In the standard CO WD
  simmering case, the reactions have been carefully worked out by
  \citet{Piro2008a}, \citet{Chamulak2008}, and \citet{Forster2010}.
  However, in the ONe WD case the \carbon[12] abundances are lower and
  the \sodium[23] / \neon[23] abundances are higher.  This means that the protons
  and $\alpha$ particles produced in carbon burning might go through
  somewhat different pathways than in the standard simmering case.}  The
development of a convective core has important consequences in the
overall evolution of the star.  Most importantly, it distributes the
entropy generated by continued carbon burning and exothermic electron
captures over a larger mass.  This has been shown to lead to oxygen
ignition at higher densities (relative to convectively stable models),
favoring collapse to an NS \citep{Miyaji1980, Miyaji1987}.

The numerical difficulties that occur once a convection zone develops
are essentially the same as those encountered in previous work, thus
we briefly recapitulate the discussion from Section 6 of
\citet{Schwab2017a}.

When using MLT in \MESA, cells that are convective have roughly the
adiabatic temperature gradient while non-convective cells have the
radiative temperature gradient.  Therefore, when cells change between
convective and radiative, they change the overall temperature profile
and in turn, this can alter the convective stability of neighboring
cells.  (This is further complicated by the fact that the
sub-threshold electron-capture reaction rates are exponentially
sensitive to temperature.)  As \MESA\ iterates to find the solution to
its equations, convective boundaries change location
iteration-to-iteration and the solver fails to converge.

It is clear that regions become convectively unstable, but an
understanding of how and whether these convective regions grow remains
an outstanding problem.  If a long-lived central convection zone does
form, matters are further complicated as this allows for the operation
of the convective Urca process \citep{Paczynski1973b}.  This continues
to be difficult to model in stellar evolution codes
\citep[e.g.,][]{Lesaffre2005b} and current versions of \MESA\ produce
physically inconsistent results during such phases
\citep{Schwab2017b}.

The choice to artificially suppress convection corresponds to the
assumption that these convective regions do not grow.
The evolution of
  our ``no convection'' models will be roughly accurate only if the
  the energy and composition changes from nuclear reactions remain
  approximately local.
In this case, the presence of carbon
leads to ignition at a factor of $\approx 2$ lower density than in the
carbon-free case.  In the case that a large convective core does form
(invalidating our local treatment),
it seems clear that this leads to higher density oxygen ignition than
in a case without a convective core; in the carbon-free models of
\citet{Gutierrez1996}, non-convective models ignite at
$\logRhoc \approx 10$ with ignition in convective cores occurring at a
central density a factor of $\approx 2$ higher.  However, the presence
of carbon is likely to reduce the density relative to a carbon-free
model, as the additional heating source means that fewer electron
captures are required.  Since we cannot follow the models through a
convective Urca phase, we are unable to reliably estimate the size of
this shift.

\section{Conclusions}
\label{sec:conclusions}

We have explored the role that residual \carbon[12] present in ONe WDs
can play during their evolution towards AIC.  We used \MESA\ to
produce 12 ONe WD models with chemical abundance profiles
self-consistently generated via stellar evolution models of SAGB
stars.  We explored the range of \carbon[12] expected for different
initial masses and showed its dependence on assumptions about rotation
and convective overshooting
(Figure~\ref{fig:comparing_central_carbon}).  Our results were
consistent with previous work preformed using an independent stellar
evolution code \citep{Siess2007A&A}.

While the carbon profiles at the time of WD formation were complex and
varied, we showed that after $\approx \unit[100]{Myr}$ of cooling the
individual interior chemical profiles become nearly spatially uniform
due to the effects of convection and thermohaline mixing.  We
typically expect WDs on the way to AIC to cool post-formation, as time
is required for their binary companion to evolve and donate sufficient
material.
Post-cooling, a significant spread in central carbon abundances
remains, with \carbon[12] mass fractions ranging from 0.1 to 10 per
cent (Figure~\ref{fig:comparing_central_carbon_post_cool}).

We then construct simple homogeneous ONe WD models with varying carbon
fractions and allow them to grow via accretion.
This allows us to map out the effects of the \carbon[12] in a
controlled way. In contrast to \citet{Gutierrez2005}, we did not find
the presence of low-density carbon ignition.  Instead, we find that
the occurrence of Urca-process cooling implies that models are
unlikely to reach carbon ignition before \magnesium[24] electron
captures begin around $\logRhoc \approx 9.6$.  We find that models
with very low carbon abundances (mass fraction $\la 0.003$) evolve
similarly to carbon-free models whereas models with higher carbon
abundances (mass fraction $\ga 0.01$) ignite carbon burning during the
electron captures.  This change in behavior happens within the range
of carbon fractions found in our more realistic ONe WD models.

Ultimately, the goal of studies such as this one is to advance our
understanding of the final fates of accreting ONe WDs that approach
the Chandrasekhar mass.  A key indicator of the likely outcome is the
central density at the time of thermonuclear oxygen ignition.  The
exact line between explosion and collapse has not been firmly
delineated, but seems to lie around $\logRhoc \approx 9.9$
\citep[e.g.,][]{Timmes1992b, Leung2017}.  In the case of explosion,
the models of \citet{Jones2016} leave behind a low mass bound remnant.
Intriguingly, several WDs with low masses and peculiar compositions
have recently been discovered \citep{Gansicke2010, Kepler2016b,
  Vennes2017a, Raddi2018}.  Understanding the connection between these
objects and AIC is an important motivator in understanding the
boundary between explosion and collapse.

Difficulties associated with modeling convective regions in these
degenerate interiors prevent us from evolving our \MESA\ models all
the way to oxygen ignition. On one hand, the presence of carbon
provides an additional energy source that, once ignited at the
lower densities associated with the \magnesium[24] electron captures,
can help heat the WD to oxygen ignition conditions.  On the other
hand, this energy release also seems likely to push the model towards
convective instability and the development of a convective core can defer
oxygen ignition to higher densities.  Firm conclusions await improved
modeling techniques.

Nonetheless, this work indicates that the inclusion of carbon burning
alters the evolution from that previously described in the literature.
Intriguingly, the range of carbon abundances seen in our WD models
produce evolutionary variations of a magnitude such that one might
expect them to lead to a diversity of outcomes.  Including the effects
of residual \carbon[12] in ONe WDs, and more generally moving towards
increasingly realistic progenitor modeling, is an important ingredient
in modeling of AIC going forward.

\acknowledgments

We thank Jared Brooks, Lars Bildsten, and Eliot Quataert for their
involvement in early explorations in this vein.  We thank them, Ken
Shen, Rob Farmer, and Frank Timmes for helpful discussions.
We thank the referee for useful comments.
We thank Enrico Ramirez-Ruiz for his co-supervision of KR.
KR~also thanks the Koret Scholars program and the Lamat REU program
for financial support.
Support for this work was provided by NASA through
Hubble Fellowship grant \# HST-HF2-51382.001-A awarded by the Space
Telescope Science Institute, which is operated by the Association of
Universities for Research in Astronomy, Inc., for NASA, under contract
NAS5-26555.  This research made extensive use of NASA's Astrophysics
Data System.


\software{\MESA\ \citep{Paxton2011, Paxton2013, Paxton2015, Paxton2018},
  \texttt{Python} (available from \href{https://www.python.org}{python.org}),
  \texttt{matplotlib} \citep{matplotlib},
  \texttt{NumPy} \citep{numpy},
  \texttt{py\_mesa\_reader} \citep{pmr}
}

\bibliographystyle{aasjournal}
\bibliography{aic_with_carbon}

\end{document}